\documentclass[pre,final,showpacs,twocolumn,amsmath,amssymb,superscriptaddress]{revtex4-1}
\usepackage[utf8]{inputenc}
\usepackage{amsmath}
\usepackage{amsfonts}
\usepackage{amssymb}
\usepackage{graphicx}
\usepackage{color}

\newcommand{\rma}{{\mathrm{a}}}
\newcommand{\rmd}{{\mathrm{d}}}
\newcommand{\rme}{{\mathrm{e}}}
\newcommand{\rmg}{{\mathrm{g}}}

\newcommand{\rmAr}{{\mathrm{Ar}}}
\newcommand{\rmRF}{{\mathrm{RF}}}
\newcommand{\rmD}{{\mathrm{2D}}}
\newcommand{\lamcr}{\lambda_{\mathrm{cr}}}
\newcommand{\rmMZ}{{\mathrm{MZ}}}
\newcommand{\rmon}{{\mathrm{on}}}
\newcommand{\rmth}{{\mathrm{th}}}
\newcommand{\rmtr}{{\mathrm{tr}}}
\newcommand{\rmt}{{\mathrm{t}}}

\begin{document}

\title{Thermal runaway in two-dimensional complex plasma crystals}

\author{L. Cou\"edel}
\email{lenaic.couedel@univ-amu.fr}
\affiliation{CNRS, Aix Marseille Univ., PIIM, UMR 7345, 13397 Marseille cedex 20, France.}
\affiliation{Department of Physics, Auburn University, Auburn, Alabama 36849, USA}

\author{V.~Nosenko}
\address{Institut f\"ur Materialphysik im Weltraum, Deutsches Zentrum f\"ur
Luft und Raumfahrt, D-82234 We{\ss}ling, Germany}
\author{M.~Rubin-Zuzic}
\address{Institut f\"ur Materialphysik im Weltraum, Deutsches Zentrum f\"ur
Luft und Raumfahrt, D-82234 We{\ss}ling, Germany}
\author{S.~Zhdanov}
\address{Institut f\"ur Materialphysik im Weltraum, Deutsches Zentrum f\"ur
Luft und Raumfahrt, D-82234 We{\ss}ling, Germany}
\author{Y. Elskens}
\affiliation{CNRS, Aix Marseille Univ., PIIM, UMR 7345, 13397 Marseille cedex 20, France.}
\author{T. Hall}
\affiliation{Department of Physics, Auburn University, Auburn, Alabama 36849, USA}
\author{A. V. Ivlev}
\affiliation{Max Planck Institute for extraterrestrial Physics, D-85741 Garching, Germany}

\date{\today}

\begin{abstract}
The full melting of a two-dimensional plasma crystal was induced in a principally stable monolayer  by localized laser stimulation. 
Two distinct behaviors of the crystal after laser stimulation were observed depending on the amount of injected energy: 
(i) below a well-defined threshold, the laser melted area recrystallized; (ii) above the threshold, it expanded outwards in a 
similar fashion to  mode-coupling instability induced melting, \textcolor{black}{rapidly destroying the crystalline order of the whole complex plasma monolayer}. 
The reported experimental observations are due to 
the fluid mode-coupling instability which can pump energy into the particle monolayer at a rate surpassing the heat transport 
and damping rates in the energetic localized melted spot, resulting in its further growth. 
This behavior exhibits remarkable similarities with impulsive spot heating and thermal runaway (explosion) in 
ordinary reactive matter. 
\end{abstract}


\maketitle 

\section{Introduction}

Complex (or dusty) plasmas are weakly ionized gases containing solid micro- or 
nanoparticles (often referred to as dust particles). 
Due to interactions with electrons and ions in the ambient plasma, dust accumulates a net electric charge \cite{Bouchoule1999, Shukla2002,Morfill2009}. 
In laboratory low-pressure gas discharges, this charge is negative due to 
the high mobility of the electrons. By injecting monosized spherical microparticles in the sheath of a capacitively coupled radio-frequency 
(cc-rf) discharge, it is possible for particles to form a monolayer. This is a result of all the microparticles levitating at the same height 
in the sheath region above the lower electrode where the electric field is strong enough to balance gravity and ensure stiff 
confinement \cite{Brattli1996,Samarian2001}. Under  specific conditions, they  form strongly coupled 
 ordered structures known as two-dimensional (2D) complex plasma crystals \cite{Dubin2000,Melzer2000,Nunomura2000,Samsonov2000,Chu1994,Thomas1994}. 

2D complex plasmas are often used as  model systems for studies of 
generic phenomena occurring in liquids and crystals at the kinetic (particle) level \cite{Morfill2009}. In such systems, it is indeed possible to obtain 
complete information about the state of the whole system of particles in the kinetic 
$(\mathbf{x},\mathbf{v})$-space. This offers an important advantage for the investigation 
of collective processes occurring in strongly coupled media including melting  and recrystallization \cite{Schweigert1998,Melzer1996,Melzer1996a,Thomas1996},
mass and heat diffusion \cite{Nosenko2008}, solitons and shocks \cite{Samsonov2005}.

In 2D complex plasmas, as in any strongly coupled 2D
system, two in-plane wave modes can be sustained. In crystals, one of these modes is longitudinal, while the other is transverse, 
with both having an acoustic dispersion \cite{Qiao2003}. Since the strength of the
vertical confinement due to the sheath electric field is finite, there is a
third fundamental wave mode with negative optical dispersion associated with the out-of-plane
oscillations \cite{Liu2003b,Couedel2009a}. The propagation of dust-lattice (DL) waves in these 2D lattices is often used as a diagnostic tool to
determine parameters of the plasma crystal since the wave dispersion 
depends directly on them \cite{Nunomura2002,Nunomura2000,Nunomura2003,Nunomura2005}.

In the sheath of a cc-rf discharge, the ion stream coming from the bulk plasma 
and directed toward the electrode is   focused
downstream of each negatively charged particle of the monolayer and  creates a perturbed region 
called the ``ion (or plasma) wake''. Ion wakes exert an attractive force on the neighboring particles.
They can thus be considered as an (external) ``third body''
in the interparticle interaction resulting in non-reciprocal particle 
pair interactions \cite{Ishihara1997,Nunomura1999,Melzer2000a,Hebner2004}.
Under certain conditions, this can lead to
the formation of an unstable hybrid mode and the melting
of the crystalline monolayer when the out-of-plane mode crosses the in-plane 
compressional mode: the mode coupling instability (MCI)
\cite{Ivlev2001,Zhdanov2009,Couedel2011,Ivlev2014,Roecker2014}. The hybrid mode has
clear fingerprints: critical angular dependence, a mixed polarization, 
distinct thresholds \cite{Couedel2011}, and synchronization
of the particle motion \cite{Couedel2014a}.

Wake-induced mode coupling is also possible  in  liquid complex plasma monolayers \cite{Ivlev2014}. 
Remarkably, in such cases, 
the confinement and dust particle
density thresholds, which are important features of the MCI 
in 2D complex plasma crystals, disappear and
the instability has a higher growth rate.  Consequently, conditions exist for which both the crystalline and the fluid 
states are viable, meaning no crossing of the modes in the crystal state and MCI growth rate high enough in 
the fluid state to prevent crystallization. This  explains why, in many experiments, the crystallization of a 
monolayer is achieved only by increasing the gas pressure and/or the rf power (confinement strength is an 
increasing function of the discharge power \cite{Steinberg2001}). 
In addition, it is in principle possible to trigger sporadic melting of a stable crystal which is not too far from the crystalline MCI threshold by 
applying a sufficiently strong mechanical
perturbation.  

In this article, we demonstrate experimentally  
that wake-mediated resonant mode coupling can be induced in a principally stable  2D
plasma crystal (no crossing of the modes) levitating in the sheath of a 
cc-rf discharge through an external  excitation mechanism. 
Localized laser stimulation of the monolayer can trigger MCI-induced melting of the 
stable crystal if the injected energy is sufficient to 
create a melted spot in which  
the excess energy due to the fluid MCI is high enough to overcome neutral damping and heat 
conduction in the rest of the crystal. This causes a further increase in 
temperature leading to the extension of the melted area over the whole monolayer in a kind of uncontrolled 
 positive feedback. 
This ``explosive behavior'' exhibits similarities with impulsive spot 
heating and thermal runaway in ordinary matter. 

\section{Experimental setup}

\begin{figure}[htbp]
	\centering
	\includegraphics[width=\columnwidth]{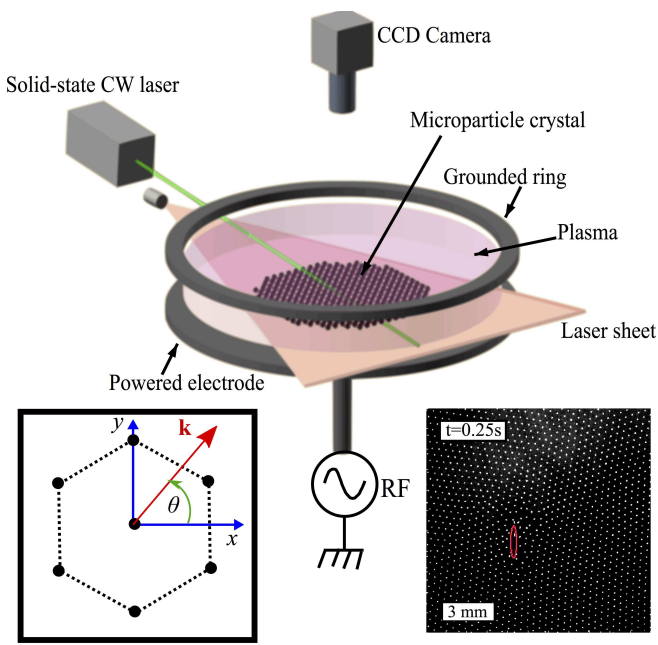}
	\caption{(Color online) Sketch of the experimental setup.
	The left inset shows an elementary cell of the hexagonal lattice 
	and the frame of reference chosen in this article. The 
	orientation of the wave vector $\mathbf{k}$ is measured in respect to the $x$-axis.
	The right inset is a snapshot of the crystal after 0.25~s of laser heating. The red ellipse shows where the 
	laser light beam interacts with the monolayer.}\label{fig:setup}
\end{figure}
The experiments were carried out in a  cc-rf glow discharge at 13.56 MHz (modified GEC chamber). A 
sketch of the setup is presented in Fig.~\ref{fig:setup}.
The argon pressure $p_{\rmAr}$ was between 0.6~Pa and 1~Pa, and the forward
rf power $P_{\rmRF}$ was between 10 W and 20 W.
In the bulk discharge, the electron temperature was
$T_{\rme}=2.5$~eV and the electron density  was $n_{\mathrm{e}}=2 \times10^9$~cm$^{-3}$
at $p_{\mathrm{Ar}}=$0.66~Pa and $P_{\rmRF}$=20~W \cite{Nosenko2009}.
Melamine-formaldehyde spherical particles with a diameter of $9.19\pm0.09~\mu$m were levitated in the plasma sheath above the lower rf
electrode and formed a horizontal monolayer, up to 60~mm in diameter.
The dust particle suspension was illuminated by a horizontal laser sheet.
The particles were imaged through a window at the top of the chamber by a 4 Megapixel
Photron FASTCAM Mini WX100 camera at a
speed of 250 frames per second. The particle horizontal coordinates, 
$x$ and $y$, and velocities, $v_x$ and $v_y$, were then extracted with 
sub-pixel resolution in each frame. An additional side-view
camera (Basler Ace ACA640-100GM) was used to confirm that we were indeed working 
with a single layer of particles. More details can be found in previous publications
\cite{Couedel2011,Couedel2012, Roecker2014}.

A SpectraPhysics Millennia PRO 15sJ solid-state 532-nm laser with a continuous-wave (cw) output power of 0.30-2~W was focused 
for a short period of time on the crystal. Particles in the laser spot were accelerated and the crystal was melted locally. 
At the crystal surface, the laser spot had an elliptical shape with its major axis in the direction of propagation and the 
illuminated area was $S_{\mathrm{L}}\simeq0.32$~mm$^2$. The duration of the laser pulse was varied from 0.05~s to 0.65~s.  

\begin{figure}[htbp]
	\centering
	\includegraphics[width=\columnwidth]{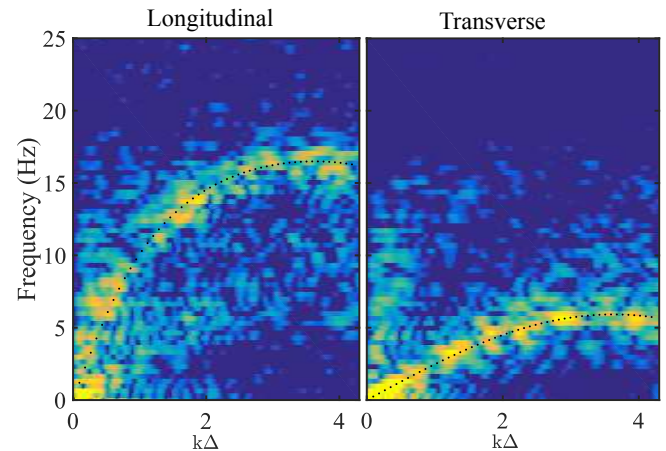}
	\caption{(Color online) Longitudinal and transverse current fluctuation spectra at an angle $\theta=0^{\circ}$. The dotted black 
	curves represent the fitted dispersion relations.}\label{fig:spectra}
\end{figure}

\section{Results}

\subsection{Characterization of the stable crystal}

Prior to laser excitation, a stable 2D crystal with respect to the mode coupling instability (no crossing of the eigenmodes 
\cite{Couedel2011}) was levitated above the rf electrode. In the set of experiments presented in this article, the following discharge conditions 
were chosen: $p_{\rmAr}=1.04$~Pa and $P_{\rmRF}=12$~W.  
For each run, the same microparticles were used and, in case of melting induced by the laser excitation, the monolayer 
was recrystallized by increasing argon pressure and rf power. The discharge conditions were then restored to the chosen values 
and the crystalline monolayer was allowed to relax a few minutes before the next laser pulse. 

From the particle positions,   the 
mean interparticle distance $\Delta=415\pm10$~$\mu$m (number density  $n_{\rm2D}=675\pm20$ cm$^{-2}$) was obtained by calculating 
 the pair correlation function and measuring the position of the first peak. Using particle velocities,
the in-plane compressional and shear particle current fluctuation spectra of the crystalline monolayer 
were computed. In Fig.~\ref{fig:spectra}, the spectra are shown for a propagation angle $\theta=0^{\circ}$ 
(see inset of Fig.~\ref{fig:setup}). No fingerprints of the unstable hybrid mode 
could be detected and only the eigenmodes were present. By fitting simultaneously the in-plane 
compressional and shear current fluctuation spectra by the theoretical curves of a 2D Yukawa crystal \cite{Zhdanov2009}, 
the microparticle charge $Q=-13800e \pm 500e$, where $e$ is the elementary charge,  
the screening length $\lambda_{\mathrm{D}}=408\pm47$~$\mu$m and the screening parameter $\kappa=1.0\pm0.15$ were obtained. 
The vertical confinement frequency was $\sim25$~Hz which is in perfect agreement with a crystal far from the MCI threshold.

\subsection{Propagation of the melting front: a threshold behavior}

\begin{table}[htbp]
\caption{\label{tab:exp_melt} Laser power $P_{\mathrm{l}}$, pulse duration $t_{\rmon}$, the sum $K_{\mathrm{t}}$ of the kinetic energy of 
the particles in the camera FoV, and, depending on whether the crystal recrystallized at the end of 
the laser pulse or melted, the transition time $t_{\mathrm{tr}}$ before recrystallization or the 
inverse of energy growth rate $\gamma_{\mathrm{g}}^{-1}$, respectively.  The boldface entries are the experiments for 
which a full melting of the monolayer was induced by the laser pulse.}
	\begin{ruledtabular}
		\begin{tabular}{cccccc}
			Exp. & $P_{\mathrm{l}}$ 	& $t_{\rmon}$ & $K_{\mathrm{t}} (t_{\rmon})$ & $K_{\mathrm{t}} (t_{\rmon}+t_{\rmth})$ & $t_{\rmtr}$, $\mathrm{\gamma_{\rmg}^{-1}}$   \\
			\# & (W)				&   (s)	   &  (keV)					& (keV)										& 			(s)			\\
			\hline
			1 & 0.3				&	0.50 	& 6.22$\pm$0.5		& 				3.1$\pm$0.1								&      		-					\\
			2 & 0.35				&	0.50		& 7.65$\pm$0.5		&				3.6$\pm$0.1								&				-					\\
			3 & 0.40				&	0.50		& 11.3$\pm$1.0		&				4.8$\pm$0.2								&				-					\\
			4 & 0.45				&	0.50		& 13.0$\pm$1.4		&				7.7$\pm$0.5								&				-					\\
			5 & 0.50				&	0.50		& 18.2$\pm$2.0		&				8.3$\pm$0.2								&				-					\\
			6 & 0.55				&	0.50		& 23.0$\pm$2.1		&				14.6$\pm$0.5								&				-					\\
			7 & 0.60				&	0.50		& 21.9$\pm$3.0		&				15.8$\pm$0.2								&				-					\\
			8 & 0.65				&	0.50		& 25.5$\pm$1.5		&				21.0$\pm$0.4								&	0.59$\pm$0.03		\\
			9 & 0.70				&	0.50		& 28.5$\pm$2.4		&				20.9$\pm$1.2								&	-		\\
			10 & 0.75				&	0.50		& 33.1$\pm$3.5		&			23.0$\pm$1.0									&	-		\\
			11 & 0.80				&	0.50		& 40.1$\pm$1.3		&			37.1$\pm$1.1									&	0.95$\pm$0.03		\\
			12 & 0.85				&	0.30		& 14.4$\pm$1.3		&			9.7$\pm$0.4									&	-		\\
			13 & 0.85				&	0.40		& 42.8$\pm$3.5		&			30.7$\pm$0.6									&	0.53$\pm$0.02		\\
			\textbf{14} & \textbf{0.85}	&	\textbf{0.50}		& \textbf{48.1$\pm$4.1}		&			\textbf{45.7$\pm$0.7}		&	\textbf{0.92$\pm$0.02}		\\			
			15 & 0.85				&	0.50		& 34.3$\pm$1.1		&			37.1$\pm$0.5									&	1.86$\pm$0.03		\\
			16 &  0.85				&	0.50		& 43.4$\pm$3.2		&			35.0$\pm$1.5									&	2.60$\pm$0.10		\\
			\textbf{17} & \textbf{0.85}				&	\textbf{0.55}		& \textbf{43.3$\pm$2.7}		&			\textbf{40.9$\pm$1.1}				&	\textbf{0.87$\pm$0.04}		\\
			\textbf{18} & \textbf{0.85}				&	\textbf{0.55}		& \textbf{52.4$\pm$2.1}		&			\textbf{63.0$\pm$1.8}				&	\textbf{1.26$\pm$0.01}		\\
			19 & 0.85				&	0.55		& 42.6$\pm$3.9		&			40.0$\pm$1.3				&	3.13$\pm$0.03		\\
			\textbf{20} & \textbf{0.85}				&	\textbf{0.55}		& \textbf{49.9$\pm$1.9}		&			\textbf{51.9$\pm$1.2}				&	\textbf{0.97$\pm$0.01}	    \\
			\textbf{21} & \textbf{0.85}				&	\textbf{0.60}		& \textbf{60.6$\pm$1.3}		&			\textbf{62.2$\pm$1.2}				&	\textbf{1.21$\pm$0.01}		\\
			\textbf{22} & \textbf{0.85}				&	\textbf{0.65}		& \textbf{64.1$\pm$14.3}	&			\textbf{58.3$\pm$10.2}			&	\textbf{1.21$\pm$0.01}		\\
			23 & 2.00				&	0.05		& 12.0$\pm$7.4		&			8.0$\pm$0.2				&	-	\\
			\textbf{24} & \textbf{2.00}				&	\textbf{0.20}		& \textbf{40.9$\pm$3.6}		&			\textbf{41.0$\pm$2.0}				&	\textbf{1.23$\pm$0.01}		\\
		\end{tabular}
	\end{ruledtabular}
\end{table}

The laser beam was focused on the crystal  so that only a few particles  were accelerated 
during the pulse ($N\leq 50$). It created, at a few $\Delta$ in front of the laser spot position in 
the direction of laser light propagation, a small melted spot  
($1-4$~mm  in diameter) containing particles having high velocities. 
Depending on the pulse duration and the laser power, the melted area recrystallized or expanded over the whole crystalline monolayer. 
By tracking the particles during and after the laser pulse, the evolution of the particle kinetic energies was measured, and the energy injected by the 
laser into the crystal was estimated. At the end of the laser pulse, the total kinetic energy of the particles in the 
camera field of view (FoV) was at its maximum. However, the particles in the melted spot had trajectories directed along the direction of the laser beam.
By looking at the first and second moments of the velocity distribution 
function in $x$ and $y$, thermalization due to collisions with other particles was completed after 
$t_{\rmth}=0.5$~s (similar Maxwellian velocity distribution \textcolor{black}{centered on 0 in $x$ and $y$ \footnote{ \textcolor{black}{$t_{\rmth}$ slightly depends 
on the laser pulse duration and the laser power; the chosen $t_{\rmth}$ corresponds to the value at which the microparticles are thermalized in all experiments.}}}).

In Table~\ref{tab:exp_melt}, the sum of the kinetic energies of 
the particles  $K_{\rmt}=\sum_i K_i(t)$ (total energy) in the camera FoV  at the 
end of the laser pulse ($K_{\mathrm{t}} (t_{\rmon})$) and 
after thermalization  ($K_{\mathrm{t}} (t_{\rmon}+t_{\rmth})$) are listed for the different runs. 
As expected, at a given laser power, the total energy increased with the pulse duration. The energy after 
thermalization was slightly smaller due to damping
by the neutral background and heat transport in the whole crystalline suspension (which was not 
entirely comprised  in the FoV). Before the laser pulse, the average kinetic temperature of the particles was estimated at 
$T_0=0.13\pm0.05$~eV, giving a total energy in the field of view $K_{\rmt}<800$~eV. Thus the measured total energy just after the laser pulse 
corresponds roughly to the energy injected by the pulse into the crystal since it was always much higher than a few keV.
\begin{figure}[htbp]
	\centering
	\includegraphics[width=0.99\columnwidth]{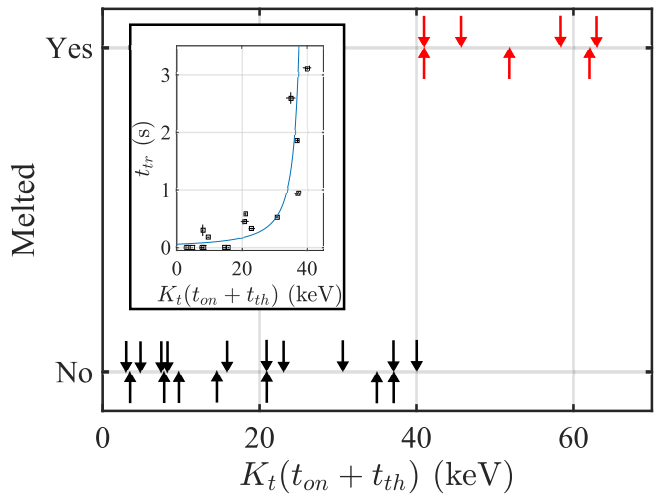}
	\caption{(Color online) State of the monolayer (red, melted or black, unmelted) at the end of the experiment as a 
	function of $K_{\mathrm{t}} (t_{\rmon}+t_{\rmth})$. Each experiment of Table~\ref{tab:exp_melt} is indicated 
 	by an arrow of the appropriate color. 
 	A threshold at $\sim$40~keV is clearly identified. The inset shows the transition 
	time before recrystalisation sets in for  injected energies below the melting threshold. The blue line is a guide for the eye. }\label{fig:transition}
\end{figure}

In Fig.~\ref{fig:transition}, the state of the monolayer at the end of the experiment (recrystallized or fully melted) 
is plotted as a function of the total energy after thermalization. 
As can be seen, when  $K_{\rmt}(t_{\rmon}+t_{\rmth})$ reached $=40\pm1$~keV (corresponding to  $t_{\rmon}\geq500$~ms at $P_{\mathrm{l}}=0.85$~W), 
the crystalline monolayer fully melted. The observed step function is a clear sign of a threshold phenomenon. 
\begin{figure}[htbp]
	\centering	
	\includegraphics[width=0.99\columnwidth]{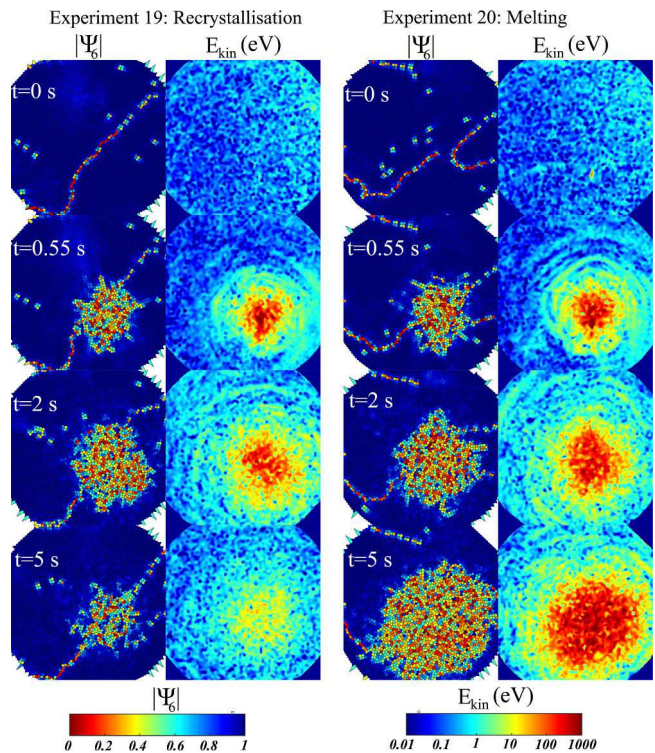}
	\caption{(Color online) $|\Psi_6|$ and energy maps of the crystal during Exps.~19~and~20 (see Table~\ref{tab:exp_melt}). Each map 
	is a square of 24.6$\times$24.6~mm$^2$.}\label{fig:snapshots}
\end{figure}

\begin{figure}[htbp]
	\centering
	\includegraphics[width=0.80\columnwidth]{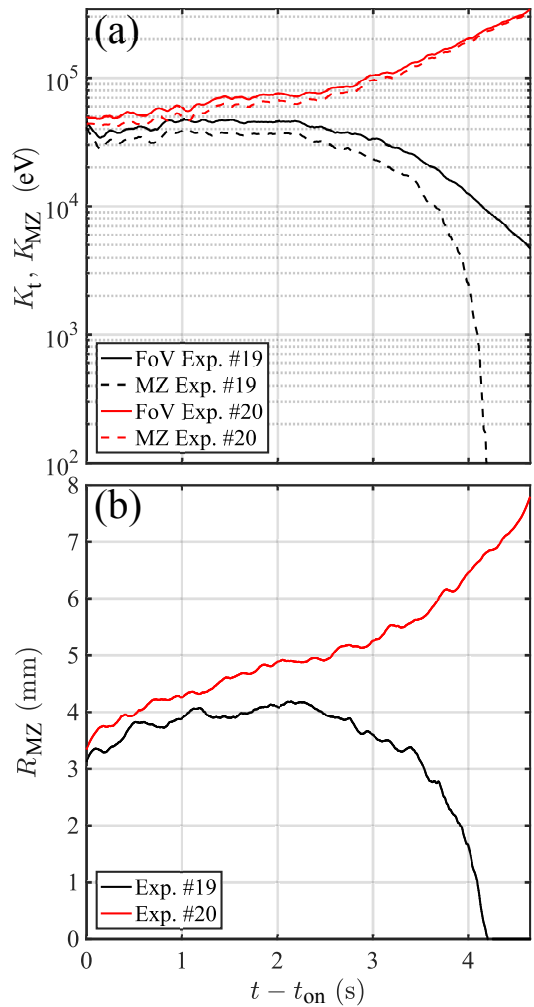}
	\caption{(Colour online) (a) Evolution of the sum the kinetic energies of the
	 particles in the field of view and in the melted zone as a 
	function of time for Exps.~19~and~20 (see Table~\ref{tab:exp_melt}). (b) Evolution of the radius of 
	the melted zone as a function of time.}\label{fig:Evolution}
\end{figure}

In Figs.~\ref{fig:snapshots} and \ref{fig:Evolution}, we illustrate the difference between two experiments showing either recrystallization 
or melting  (Exps.~19~and~20, respectively; see Table~\ref{tab:exp_melt}).
In Fig.~\ref{fig:snapshots}, the monolayer Voronoi map (colored according to $|\Psi_6|$) and the kinetic energy map are presented for 
two experiments using the same laser
pulse setting. Here, $\Psi_6(j)=(1/N)\sum_{l=1}^N e^{i\theta_{jl}}$, where $N$ is the number 
of closest neighbors and $\theta_{jl}$ is the angle of vector linking the $j$th to the $l$th particle with  respect 
to the $x$-axis. This order parameter $|\Psi_6|$ represents 
the deformation of a hexagonal cell from a perfect hexagon ($|\Psi_6|=1$). As can be seen, before the laser pulse ($t=0$~s), 
the crystalline monolayer was cold in both cases with $T_0\simeq0.13$~eV and the crystals were almost perfect with only a few dislocations and 
a few grain boundaries. At the end of the laser pulse ($t=0.55$~s),  a melted spot (high 
kinetic energy and low values of $|\Psi_6|$) could be seen in the center of 
the frame. In both cases, compressional waves traveled preferentially in the direction of the laser beam (positive $y$ direction) due 
to the non-thermalized motion of the excited particles. At \textcolor{black}{$t=2$~s $>t_{\rmon}+t_{\rmth}$},  the particles were thermalized and compressional waves  emerged  
from the hot spot isotropically. It was observed that,  in both cases, the melted spot became slightly 
larger than at $t=0.55$~s due to heat transport in the 
crystalline monolayer. However, the spot in experiment 20 showed a slightly larger melted spot and the particle kinetic energies were slightly higher.

 This observation is confirmed in Fig.~\ref{fig:Evolution}, in which  the time
evolution  (after the laser pulse) of the radius of the melted zone (MZ) and 
the evolution of the total kinetic energy in the 
MZ and the camera FoV are plotted. \textcolor{black}{In this article, the MZ is defined as the area in which the mean 
kinetic energy is greater than 30~eV \footnote{\textcolor{black}{For the measured crystal parameters, the melting occurs 
when the microparticle mean kinetic energy is around 10-20 eV \cite{Hartmann2005,Kryuchkov2017}. It slightly 
underestimates the size of the MZ but catches the main features of its evolution  
without including particles in the crystalline state}}}. As can be seen in both experiments, most of the kinetic energy in the FoV is carried by the particles in the MZ, though 
in experiment 19, the injected energy  and the radius of the MZ were slightly smaller. In both cases the kinetic energy and MZ radius initially 
increased, but in experiment 19 the total energy inside the MZ became stable, meaning that the kinetic energy began to decrease. On the 
contrary, $K_{\rmMZ}$ continued to grow in the case of experiment 20. After $\sim2.25$~s, the MZ started to shrink in experiment 19 and the total energy 
started to decrease  while both 
parameters kept growing in the case of experiment 20 (the energy growth seemed exponential). Finally, in the case of experiment 19, 
 after $\sim3.1$~s, the total energy exhibited an exponential decay while the MZ was rapidly shrinking, denoting recrystallization. 
 In experiment 20, the energy kept growing and the MZ eventually extended to the whole monolayer.  After 3.5~s, the radius 
 of the MZ grew quasi linearly with a velocity ${v}_{\rmMZ}\sim2$~mm/s. A snapshot of the monolayer 
 during  the final stage of both experiments can be seen in Fig.~\ref{fig:snapshots} ($t=5$~s).

Another interesting observation is the existence of a transition period before the recrystallization sets in for injected energy 
below the melting threshold. For the purpose of analysis, the transition time 
$t_{\mathrm{tr}}$ is defined as the time before $K_{\mathrm{t}}$ exhibits a clear exponential decay after the end 
of the laser pulse \footnote{In logarithmic scale, $t_{\mathrm{tr}}$ is the time at which 
the straight line with a negative slope fitting  the exponential decay 
of the energy  at large $t$ crossed the line defined 
by $K_{\rmt}=K_{\mathrm{t}} (t_{\rmon}+t_{\rmth})$}. In  experiment 19, the transition time was 3.13~s 
(see Fig.~\ref{fig:Evolution}a). As can be seen in the inset of Fig.~\ref{fig:transition}, the transition 
time gets longer as the input energy gets closer to the induced-melting energy threshold (experimental  values 
are listed in Table~\ref{tab:exp_melt}). For the experiments with injected energy above the 
threshold the energy growth looked nearly exponential. In 
Table~\ref{tab:exp_melt}, the measured growth times $\gamma_{\rmg}^{-1}$ are listed \footnote{$\gamma_{\rmg}$ 
is obtained by fitting the following function 
to the total energy: $K_{\rmt}(t+t_{\rmon}+t_{\rmth})=K_{\mathrm{t}} (t_{\rmon}+t_{\rmth})+A\exp(\gamma_{\rmg}t)$ 
where $A$ is a positive constant}. The average
measured value was $\gamma_{\rmg}^{-1}=1.1\pm 0.2$~s.

\section{Discussion}

As reported in the previous section, the laser-induced melting of the crystal exhibited a 
clear energy threshold: above a well-defined amount of energy injected by the laser, the full melting of the crystalline 
monolayer was triggered. The mechanism by which the monolayer melted can be understood when taking into account 
the existence of the  MCI in fluid monolayers. It was indeed shown by Ivlev \textit{et al.} \cite{Ivlev2014} that 
 the wake-induced coupling of wave modes always occurs in 2D fluid complex plasmas and, contrary to MCI onset in 2D 
 complex plasma crystals,  there is no confinement-density threshold. This means that for sufficiently low damping rates 
 the energy input from the fluid MCI (i.e. energy transferred by 
the ion flow to the crystal) can prevent crystallisation of the monolayer.  Moreover, if a large enough melted spot is induced in the 
crystalline monolayer, the energy input from the localized fluid MCI will not be dissipated rapidly enough (through neutral damping and 
heat conduction in the crystal), causing a further increase in temperature triggering the expansion of the melted area in 
an uncontrolled positive feedback \textcolor{black}{(thermal runaway)}. This mechanism will lead to the total 
 destruction of the crystalline monolayer as observed for regular MCI-induced melting of 
 2D complex plasma crystals \cite{Couedel2011,Ivlev2014,Roecker2014,Williams2012,Yurchenko2017}. 
 Such an ``explosive   melting'' of a 2D complex plasma crystal  has been recently studied in detail by 
 Yurchenko \textit{et al.}  \cite{Yurchenko2017} and was found to be  
  similar to impulsive spot heating and thermal explosion (thermal runaway) in ordinary matter \cite{Ivlev2015b}. 
 
In the presented experiments, the laser pulse created a localized melted spot in the crystal where the fluid 
MCI exists. The spatial temperature distribution $T(r,t)$ in a continuous reactive medium, including damping, can be described 
by the following  heat equation  \cite{Ivlev2015b,Yurchenko2017}
\begin{equation}\label{eq}
\frac{\partial T}{\partial t}= \frac{Q(T)}{Cn_{\rmD}}-\frac{2\gamma_{\rmd}}{C}(T-T_0)
+  \frac{\chi}{r}\frac{\partial }{\partial r} \Big( r \frac{\partial T}{\partial r} \Big) 
\end{equation}
with the initial condition
\begin{equation}
	T(r,0)=T_0+\frac{E_{\mathrm{l}}}{2 \pi w_{\rmMZ}^2 n_{\rmD}C}\exp\Big( -\frac{r^2}{2w_{\rmMZ}^2}\Big)
\end{equation}
where   $E_{\mathrm{l}}$ is the energy injected by the laser pulse, $w_{\rmMZ}$ is the width of the melted zone, 
$C\simeq2-3$  is the heat capacity per particle, $\gamma_{\rmd}$  is the Epstein damping rate (for 
our experimental conditions, 
$\gamma_{\rmd}\simeq1.21\pm0.13$~s$^{-1}$ \cite{Liu2003}), and $\chi$ is the thermal diffusivity. 

In Eq.~(\ref{eq}), the first term in the r.h.s. is the heat input due to the fluid MCI. Since  fluid MCI essentially develops 
when the correlations between the particles are largely destroyed (so that a horizontal layer, uniform and of infinite
extent, can be considered \cite{Ivlev2014}), it supposes a significant overheating compared to the 
melting temperature $T_{\mathrm{m}}$ of a 2D complex plasma crystal ($T_{\mathrm{m}}\sim10-20$~eV in our condition \cite{Hartmann2005,Kryuchkov2017}). Consequently, the MCI heat source term requires the 
introduction of an activation temperature $T_{\rma}$  well above $T_{\mathrm{m}}$. It was indeed demonstrated that pair correlations are destroyed 
when the monolayer temperature is well above the melting temperature \cite{Fortov2005a,Sheridan2009}. Accordingly, the fluid MCI heat source can be
roughly modeled as a reaction rate governed by the Arrhenius law
\begin{equation}\label{eq2a}
\frac{Q(T)}{Cn_{\rmD}}= \frac{\gamma_{\mathrm{\textsc{mci}}}  T_{\infty}}{C} \mathrm{e}^{-T_{\rma}/T}\ ,
\end{equation}
which can be approximated by \cite{Yurchenko2017}
\begin{equation}\label{eq2b}
 \frac{Q(T)}{Cn_{\rm2D}}=\begin{cases}
    0, &  T< T_{\rma},\\
    \frac{\gamma_{\textsc{mci}}  T_{\infty}}{C}, & T\geq T_{\rma},
  \end{cases}
\end{equation}
where  $\gamma_{\mathrm{\textsc{mci}}}  T_{\infty}$  is the  saturated fluid MCI heat source with 
the fluid MCI growth rate $\gamma_{\mathrm{\textsc{mci}}}$   and the saturation temperature $T_{\infty}\sim1-3$~keV. 
In a recent study on the melting front propagation in MCI-induced 2D crystal melting, the activation 
temperature was found to be $T_{\rma}\sim150$~eV  \cite{Yurchenko2017}.
The saturation temperature $T_{\infty}$ arises from the thermal spreading of the monolayer  which inhibits the mode crossing and consequently 
the fluid MCI growth rate \cite{Ivlev2014}. 

Since $T_0 \ll T_{\rma}$, we can neglect $T_0$. Eq. (\ref{eq}) can be then written in a dimensionless manner
\begin{equation}\label{eq3}
\frac{\partial \Theta}{\partial \tau}= \lambda \mathrm{e}^{-1/\Theta}-\Gamma \Theta
+  \frac{1}{\tilde{r}}\frac{\partial }{\partial \tilde{r}} \Big( \tilde{r} \frac{\partial \Theta}{\partial \tilde{r}} \Big)\ , 
\end{equation}
where $\Theta=T/T_{\rma}$, $\tilde{r}=r/r^*$ with ${r^*}^2=E_{\mathrm{l}}/n_{\rm2D}CT_{\rma}$, $\tau=t/t^*$ with $t^*={r^*}^2/\chi$. The initial condition becomes
\begin{equation}
	\Theta(\tilde{r},0)=\frac{1}{2 \pi \tilde{w}_{\rmMZ}^2 }\exp\Big( -\frac{\tilde{r}^2}{2\tilde{w}_{\rmMZ}^2}\Big)\ ,
\end{equation} 
with $\tilde{w}_{\mathrm{MZ}}   =  w_{\mathrm{MZ}} / r^* $. The 
normalization $\int_0^{2 \pi} \int_0^\infty \Theta (\tilde{r}, 0) 
\tilde{r} \rmd \tilde{r} \rmd \varphi = 1$ determines the characteristic length scale.
Thus, in its dimensionless form, the problem is characterized by two numbers
\begin{equation}
	\lambda=\frac{\gamma_{\mathrm{\textsc{mci}}}  T_{\infty} E_{\mathrm{l}}}{C^2 \chi n_{\rmD} T_{\rma}^2}
\end{equation} 
and 
\begin{equation}
	\Gamma=\frac{2\gamma_{\rmd} E_{\mathrm{l}}}{C^2 \chi n_{\rmD} T_{\rma}}= \frac{2 \gamma_\rmd}{\gamma_{\mathrm{MCI}}} \frac{T_\rma}{T_\infty} \lambda\ .
\end{equation}

Eq. (\ref{eq3}) is similar to the one describing impulsive spot heating and thermal explosion in 
ordinary matter  with the addition of a dimensionless damping coefficient $\Gamma$ \cite{Ivlev2015b,Yurchenko2017}. 
When there is no damping and the heating of a reactive medium occurs within a sufficiently 
localized spot (so it can be accurately described with a Dirac distribution), the subsequent thermal evolution 
is characterized only by $\lambda$ which identifies a bifurcation between two distinct regimes. 
When exceeding a critical threshold, the heat equation exhibits the explosive solution, i.e., a thermal runaway is triggered. For a 
2D system, the critical value is $\lamcr(\Gamma=0)=9.94$ \cite{Ivlev2015b}. The damping term increases 
the critical value: numerical estimates are, for example, $\lamcr(\Gamma=1.5)\sim17$ 
and $\lamcr(\Gamma=2.5)\sim21$. In our experimental conditions, we assume that the thermal diffusivity of the crystal  was 
$\chi\sim10$~mm$^2/$s \cite{Nosenko2008} and 
$ \gamma_{\mathrm{\textsc{mci}}}\sim\gamma_{\rmg}\sim0.9$~s$^{-1}$. By fitting the temperature 
distribution  after laser excitation with a 2D Gaussian for experiments near threshold, the 
width of the melted zone was $w_{\rmMZ}\sim1-2$~mm \footnote{According to Fig.~\ref{fig:Evolution}, the area 
defined by $R_{\rmMZ}$ includes $\sim90\%$ of the energy, giving $R_{\rmMZ}\sim 2 w_{\rmMZ}$ which is in 
agreement with the measured values}, which is noticeably smaller 
than the characteristic length $r^*\sim4.5$~mm.  These experimental parameters yield the 
value of $\Gamma\sim2-4$ and, at threshold energy, $\lambda\sim10-30$, in remarkable agreement with 
the theoretical estimate. Note that the width of 
the MZ has a significant effect on the energy threshold since at a given 
injected energy a wider spot  results in a significant portion of the energy being distributed 
over particles with a kinetic temperature  below the activation temperature. 
Moreover, for large melted spots ($w_{\rmMZ}\geq r^*$), the problem is no longer characterized by a single dimensionless number and the threshold depends on the temperature
of the melted spot and its size \cite{Ivlev2015b,Merzhanov1966}.

\begin{figure}[htbp]
	\centering
	\includegraphics[width=\columnwidth]{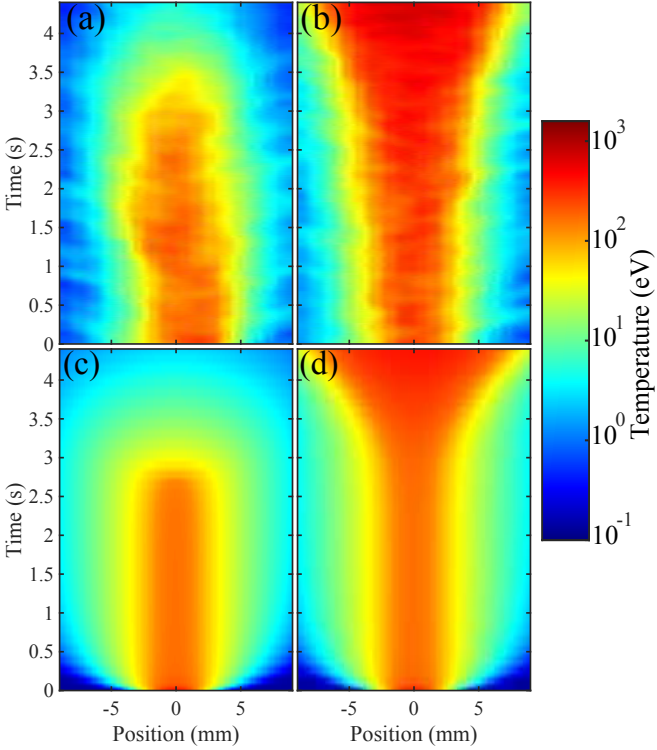}
	\caption{(Color online) Evolution of the temperature profile. (a) Experiment 19: recrystallization, (b) Experiment 20: melting. Solution of 
	Eqs.~(\ref{eq}) using the reaction model (\ref{eq2b}): (c) slightly below threshold, (d) slightly above 
	threshold. For the numerical profiles, 
	the following parameters were used: $\chi=9$~mm$^2$/s, $\gamma_{\mathrm{d}}$=1.2~s$^{-1}$, 
	$\gamma_{\mathrm{MCI}}=0.9$~s$^{-1}$, $T_{\infty}$=1800~eV, $T_{\rma}$=105~eV, 
	$n_{\rmD}$=6.75~mm$^{-2}$, $C=2$, 
	$w_{\rmMZ}$=1.15~mm and (c) $E_\mathrm{l}$=40103~eV ($\lambda=24.2497$, $\Gamma= 3.7722$) and 
	(d) $E_\mathrm{l}$=40103.7~eV ($\lambda=24.2506$, $\Gamma= 3.7722$).} \label{fig:exp_vs_th}
\end{figure}

In Fig.~\ref{fig:exp_vs_th}, the evolution of the temperature profile is presented for experiments 19 and 20 and the numerical solution of 
Eqs. (\ref{eq}) and (\ref{eq2b}). The input parameters have been chosen to reproduce qualitatively the experimental 
observations. As can be seen in all cases, the energy is transported from the MZ to the rest of the crystalline suspension.  
The instability provided energy to the microparticle suspension inside the fluid region which is partially transported in the crystal 
through heat conduction and partially dissipated  through neutral damping. For these 
reasons, the size of the melted area can slightly increase even if the total input energy is below the required melting threshold. 
As  reported in the previous section, when increasing the amount of injected energy 
while staying below threshold (for example, increasing the duration of the 
laser pulse), the crystallization is delayed. This behavior was also observed  in the 
numerical solutions and can be explained by the amount of energy pumped by the 
fluid MCI into the melted spot which depends directly on its size. Indeed,  crystallization 
can occur only after enough energy has been removed from 
the MZ. Consequently, this suppresses the fluid MCI heat flux (Fig.~\ref{fig:exp_vs_th}a and c). This transition period is longer 
for larger spots since a 
significantly larger amount of energy has to be removed due to the fluid MCI energy input. When the injected energy is slightly 
above the threshold, 
fluid MCI delivers more energy to the monolayer than what can be extracted away from the MZ 
which  grows as a result (Fig.~\ref{fig:exp_vs_th}b and d). The melting pattern is then 
similar to the one observed in the regular (crystal) MCI induced  melting of a crystal \cite{Couedel2011,Ivlev2014,Roecker2014,Zhdanov2009,Yurchenko2017}. 

In all experiments in which the melting was successfully induced, the MZ radius grew at a constant velocity. For $\tilde{r}$ large enough in Eq.~(\ref{eq3}), the last 
term of the r.h.s. tends asymptotically towards $\partial^2 \Theta/\partial \tilde{r}^2$, reducing the problem to one dimension.  
The reaction model (\ref{eq2a}) or (\ref{eq2b}) in the evolution equation (\ref{eq}) defines a bistable system, with 
two uniform stationary solutions (in the reduced version with $T_0 = 0$ and the rate equation (4), these states are 0 
and $\Theta_{\mathrm{max}} = \lambda / \Gamma$). 
With inhomogeneous boundary conditions, the front profile is a heteroclinic connection between these two states, propagating at constant velocity $v_{\mathrm{MZ}}$.
As the MZ radius grows 
linearly, a self-similar solution $\Theta(\tilde{r},\tau)=\Theta(X)$, with $X=\tilde{r}-\tilde{v}_{\rmMZ}\tau$, where $\tilde{v}_{\rmMZ}$ is the dimensionless velocity of the melting front, 
can be obtained analytically using the approximation (\ref{eq2b}) \cite{Yurchenko2017}
\begin{equation}\label{eq:sol}
 \Theta(X)=\begin{cases}
    \exp(-\sigma_1 X), &  X>0,\\
    \Theta_{\mathrm{max}}+(1-\Theta_{\mathrm{max}})\cdot \exp(-\sigma_2 X), & X\leq 0,
  \end{cases}
\end{equation}
with  $\sigma_1=(\tilde{v}_{\rmMZ}/2)+\sqrt{4\Gamma+\tilde{v}_{\rmMZ}^2}/2>0$ and  
$\sigma_2=(\tilde{v}_{\rmMZ}/2)-\sqrt{4\Gamma+\tilde{v}_{\rmMZ}^2}/2<0$. The boundary conditions 
are $\Theta(+\infty)=0$ and $\Theta(-\infty)=\Theta_{\mathrm{max}}$. The position of the melting front is given by 
$\Theta(0)=1$ and its velocity is $\tilde{v}_{\rmMZ}=-\partial_{\tau}\Theta/\partial_{X}\Theta$. It exists only if the 
two partial derivatives are continuous, giving the condition
\begin{equation}
	(1-\Theta_{\mathrm{max}}) \sigma_2=\sigma_1\ .
\end{equation} 
The velocity of the melting front is then straightforwardly obtained
\begin{equation}
	\tilde{v}_{\rmMZ}= \frac{\lambda-2\Gamma}{\sqrt{\lambda-\Gamma}}\ .
\end{equation}
For $\lambda > 2 \Gamma$ (viz.\ $\Theta_{\mathrm{max}} > 2$), the front velocity is positive, which corresponds to the melting of 
the monolayer, due to the fact that the final temperature is further away from the threshold $T_\rma$ than the cold 
crystal is ($T_\infty - T_\rma > T_\rma - T_0 = T_\rma$). For $\Gamma < \lambda < 2 \Gamma$ (viz.\ $1 < \Theta_{\mathrm{max}} < 2$), 
the front velocity is negative, corresponding to recrystallization due to the fact that the melted state equilibrium temperature is closer 
to the activation threshold than the cold crystal is. 
The case $\lambda = 2 \Gamma$ is the bifurcation case, for which the front would be stationary. It would be interesting to further investigate 
this range, recalling that model (\ref{eq2b}) is very crude and does not yield accurate values for the actual front velocity, though being 
qualitatively relevant. Under our experimental conditions,  $\tilde{v}_{\rmMZ}\sim0.5-5$ or ${v}_{\rmMZ}\sim1-10$~mm/s, which is 
in agreement with our measured value. This velocity yields the spatial decay exponents 
for the front, $\sigma_1 = (\lambda - \Gamma) / \sqrt{\lambda - \Gamma}$ 
and $\sigma_2 = \tilde{v}_{\rmMZ} - \sigma_1 =  -\Gamma / \sqrt{\lambda - \Gamma}$. The slope of the front and the 
temperature difference $\Theta_{\mathrm{max}}$ yield an estimate 
$\Theta_{\mathrm{max}}/\sigma_1 = \lambda / (\Gamma \sqrt{\lambda - \Gamma})$ for its width. Given the experimental ranges of 
$\lambda$ and $\Gamma$, the front width is $\Theta_{\mathrm{max}}/\sigma_1 \sim 1-3$ or in physical units $4.5-13.5$~mm. This value 
is close to the one observed in Figs.~\ref{fig:snapshots} and \ref{fig:exp_vs_th}.

Finally, in the frame of our model, the evolution of the total kinetic energy in the monolayer $K_{\mathrm{tot}}$ can be obtained
\begin{equation}
K_{\mathrm{tot}}(t)=2\pi n_{\rm2D} \int_0^{\infty} T(r,t) r \mathrm{d}r\ .
\end{equation}
Multiplying Eq.~(\ref{eq}) by $n_{\rm2D}$ and integrating over space,  one obtains
\begin{equation}\label{eq:J}
	\dot{K}_{\mathrm{tot}}+\frac{2 \gamma_\rmd}{C}  K_{\mathrm{tot}}= S
\end{equation}
where the heat source $S$ due to the fluid mode-coupling instability is (using the reaction model  (\ref{eq2b}))
\begin{eqnarray}
	S & = &2\pi \int_0^{\infty} \frac{Q(T)}{C} r \mathrm{d}r 
	\simeq 2\pi n_{\rm2D} \frac{\gamma_{\textsc{mci}}  T_{\infty}}{C} \int_0^{r_{\mathrm{a}}(t)}  r \mathrm{d}r \nonumber \\
	& \simeq & \pi n_{\rm2D} \frac{\gamma_{\textsc{mci}} T_{\infty}}{C}r_{\mathrm{a}}^2(t)= \pi n_{\rm2D} \frac{\gamma_{\textsc{mci}} T_{\infty}}{C}v_{\mathrm{MZ}}^2t^2\ ,
\end{eqnarray}
where $r_{\mathrm{a}}(t)$ is the front radius, such that $T(r_{\mathrm{a}},t)=T_{\rma}$. The solution of Eq.~(\ref{eq:J}) is trivial and one  immediately 
sees that, in the limit $t\rightarrow \infty$, $K_{\mathrm{tot}} \propto t^2$. This result cannot be confirmed at 
the present time by  experimental observations since the video recordings of the laser-induced melting were not long enough. In addition, 
 using the reaction model  (\ref{eq2b}), the stage of exponential growth reported in the experiments is lost. It has indeed been reported 
that, during MCI-induced melting in crystals, the microparticles kinetic 
temperature in the MZ grows exponentially until it saturates \cite{Roecker2014}. Moreover, in the 
present study, the total kinetic energy growth in the FoV seems to be nearly exponential (even though the videos were not long enough to 
recover the kinetic energy evolution over a significant energy interval to draw a firm conclusion). However,
using the reaction model (\ref{eq2a}), the stage of exponential growth can be recovered. Indeed, 
the ``source term'' in (\ref{eq3}) is
\begin{equation}\label{source}
 \Sigma(\Theta)=\lambda \exp(-1/\Theta) -\Gamma \Theta\ .
\end{equation}
For $\lambda>\Gamma \exp(1)$ (which is always the case for $\lambda>\lambda_{\mathrm{cr}}$), the source term $\Sigma$ has two nonzero real roots: the 
unstable one $\Theta_1=-1/\mathrm{W}_{-1}(-\Gamma/\lambda)<1$ and the stable one 
$\Theta_2=-1/\mathrm{W}_{0}(-\Gamma/\lambda)>1$ where $\mathrm{W}_{k}$ is the 
$k^{\mathrm{th}}$ branch of the Lambert W function. The maximum of $\Sigma$ occurs at
$\Theta_{\mathrm{gm}}=-1/(2\mathrm{W}_{\mathrm{0}}(-0.5\sqrt{\Gamma/\lambda}))$ such 
that $\Theta_1<\Theta_{\mathrm{gm}}<\Theta_2$. 
Since $ \Sigma'(\Theta_1)>0$, the stage of exponential growth can be explained by the lower, unstable root. In the early stage of the instability, 
the dimensionless heat equation can be approximated by
\begin{equation}
	\frac{\partial \Theta}{\partial \tau} -  \frac{1}{\tilde{r}}\frac{\partial }
	{\partial \tilde{r}} \Big( \tilde{r} \frac{\partial \Theta}{\partial \tilde{r}} \Big) \simeq  \Sigma'(\Theta_1)\cdot (\Theta -\Theta_1)\ .
\end{equation}
Given the experimental ranges of $\lambda$ and $\Gamma$, the unstable root is $0.28 <\Theta_1< 0.34$ and the instability 
growth rate is $5 < \Sigma'(\Theta_1)< 8$. Using our experimental parameters, the time scale is $t^*=2-3$~s giving a growth 
rate $1.7\ \mathrm{s}^{-1}<\gamma_{\mathrm{g}}<4\ \mathrm{s}^{-1}$. This range slightly overestimates the reported experimental 
value but, given the simplicity of our reaction rate model, it is in remarkable agreement. It nevertheless calls for more investigations and better modeling 
of the fluid MCI heat source in order to obtain the full physical picture of the energy input due to the fluid MCI.

\section{Conclusion}

To conclude, we have experimentally demonstrated that  wake-mediated resonant 
mode coupling  can be induced in a 2D plasma crystal  through a  
localized laser heating. This heating can trigger a thermal runaway resulting in the full melting of the crystalline monolayer.  
In a stable crystal (with respect to crystalline MCI), an energy threshold  
was observed.  Below the threshold, recrystallization was always observed with a transition period 
which grew longer when approaching the threshold. The localized fluid MCI  could not 
pump enough energy to the microparticle suspension to maintain the fluid state 
and propagate through the crystal. Above the threshold, the full melting of the crystal was observed due 
to the activated fluid MCI (\textcolor{black}{i.e. the crystalline order of the whole complex plasma monolayer was rapidly destroyed}). 
Remarkable similarities with impulsive spot heating and thermal runaway in ordinary reactive matter were reported. 

In future investigations, detailed studies of the threshold for 
various crystal parameters and laser spot sizes will be performed. The influence of the temperature of the fluid state on the MCI growth rate 
will be of particular interest.

\acknowledgments
This work was partially supported by the French-German
PHC PROCOPE Program (Project No.~35325NA/57211784). The authors would like to thank 
Prof.~E.~Thomas~Jr and Dr~I.~Laut for fruitful discussions and suggestions.

\bibliography{Biblio}

\end{document}